\documentclass[square]{ws-procs11x85}
\usepackage{multicol,float} %necessary when used with multicols
\usepackage{graphicx}
\usepackage{epsfig}

\newcommand{\massint}{k}
\newcommand{\enint}{\mu}
\newcommand{\aj}{AJ}
\newcommand{\apj}{ApJ}
\newcommand{\apjs}{ApJS}

\newcommand{\mnras}{MNRAS}

\begin{document}

\title{Magnetic acceleration of ultra-relativistic GRB and AGN jets}

\author{M. V. BARKOV$^{1,2}$ AND S. S. KOMISSAROV,$^{1}$ }

\address{$^{1}$Department of Applied Mathematics, The University of Leeds,\\
Leeds, LS2 9GT, UK\\
$^{2}$ Space Research Institute, 84/32 Profsoyuznaya Street, \\
Moscow, 117997, Russia\\
E-mail: bmv@maths.leeds.ac.uk (MVB) and serguei@maths.leeds.ac.uk (SSK)}

\begin{abstract}
We present numerical simulations of cold, axisymmetric, magnetically driven
relativistic outflows. The outflows are initially sub-Alfv\'enic and
Poynting flux-dominated, with total--to--rest-mass energy flux ratio
up to $\mu \sim 620$. 
To study the magnetic acceleration of jets we simulate flows confined within a 
funnel with rigid wall of prescribed shape, which we take to be
$z\propto r^a$ (in cylindrical coordinates, with $a$ ranging from 1 to
2). This allows us to eliminate the numerical dissipative effects induced
by a free boundary with an ambient medium.   
We find that in all cases they converge to a steady
state characterized by a spatially extended acceleration region. For the jet 
solutions the acceleration process is very efficient - 
on the outermost scale of the simulation more than half of the Poynting flux has 
been converted into kinetic energy flux, and the terminal Lorentz factor
approached its maximum possible value ($\Gamma_\infty \simeq \mu$).  
The acceleration is accompanied by the collimation of magnetic field lines in 
excess of that dictated by the funnel shape. The numerical solutions are generally 
consistent with the semi-analytic self-similar jets solutions and the spatially extended
acceleration observed in some astrophysical relativistic jets. In agreement with previous 
studies we  also find that the acceleration is significantly less effective for wind solutions 
suggesting that pulsar winds may remain Poynting dominated when they reach the 
termination shock.  
\end{abstract}


\bodymatter

\begin{multicols}{2}

%%%%%%%%%%%%%%%%%%%%%%%%%%%%%%%%%%%%%%%%%%%%%%%%%%
\section{Introduction}
%%%%%%%%%%%%%%%%%%%%%%%%%%%%%%%%%%%%%%%%%%%%%%%%%% 
\label{introduction}

There is strong evidence for relativistic motions in jets that emanate
from active galactic nuclei (AGNs). The Lorentz factors of blazar jets 
lie in the range $\sim 5-40$ \cite{J01,J05,C07}.

In the case of AGNs there have indeed been indications from a growing
body of data that the associated relativistic jets undergo the bulk of
their acceleration on scales that are of the order of those probed by
very-long-baseline radio interferometry. In particular, the
absence of bulk-Comptonization spectral signatures in blazars has been
used to infer that jet Lorentz factors $\geqslant 10$ are only attained on
scales $\geqslant 10^{17}\ {\rm cm}$ \cite{S05}.

The first theoretical clues to the necessity of relativistic motion in GRB's
arose from the compactness problem \cite{rud75}. The requirements that the 
source be optically thin can be used to obtain direct 
limits on the minimal Lorentz factor, $\Gamma \gtrsim 100$ 
\cite{kp91,pir99}. Recently, superluminal expansion was observed in the 
afterglow emission of GRB 030329 \cite{tay04}.

The main source of power of AGN and GRB jets is the rotational energy of the
central black hole \cite{L76,BZ77} and/or its
accretion disk. The naturally occurring low mass
density and hence high magnetization of black-hole magnetospheres
suggests that the relativistic jets originate directly from the
black-hole ergosphere as Poynting-dominated outflows, 
whereas the disk surface launches a slower,
possibly non-relativistic wind that surrounds and confines the highly
relativistic flow. Although, dissipative processes may directly transfer 
the electromagnetic energy to emitting particles, the commonly held view 
is that it is first converted into the bulk kinetic energy and only 
subsequently channeled into radiation through shocks and other dissipative 
waves \cite{BR74,BBR84,VK03a,VK03b,bn06, kbvk2007}. In the limit of ideal MHD 
such conversion can be achieved only via magnetic forces and  
the efficiency of this mechanism is the main subject of our investigation.

Since most of the acceleration takes place far away from the source the space-time 
is basically flat. We also use an isentropic equation of state 
$  p=Q\rho^s$, where $Q=$const and $s=4/3$. Since we are interested in the magnetic
acceleration of cold flows, we make $Q$ very small, so the gas pressure
is never a dynamical factor. This relation enables us to exclude the
energy equation from the integrated system and overcome the stiffness of relativistic
MHD in magnetically dominated regime.  

Given the condition of axisymmetry the  
poloidal magnetic field is fully described by the
so-called magnetic flux function $\Psi(r,z)$, the total magnetic flux enclosed
by the axisymmetric loop circle $r,z=$const. Moreover, for stationary flows 
there are 5 quantities that propagate unchanged along the magnetic field lines and
thus are functions of $\Psi$ alone. These are $\massint$, the
rest-mass energy flux per unit magnetic flux; $\Omega$, the angular
velocity of magnetic field lines; $l$, the total angular momentum
flux per unit rest-mass energy flux; $\enint$, the total energy flux
per unit rest-mass energy flux; and $Q$, the entropy per particle.
For cold flows ($Q=0$, $w=\rho c^2$) we have
$  \massint = {\rho \Gamma v_p}/{B_p} $, 
$  \Omega r=v^{\hat{\phi}}-v_pB^{\hat{\phi}}/B_p $, 
$    l = -I/{2\pi\massint c}+r u^{\hat\phi}$,
and
$    \enint = \Gamma \left(1+\sigma\right),$
where $\Gamma$ is the Lorentz factor,  $v_p$ is the magnitude of the poloidal velocity, 
$B_p$ and $B^{\hat{\phi}}$ are the magnitudes of poloidal and azimuthal magnetic field,  
$ I = c r B^{\hat\phi}/2 $ is the total electric current flowing through a loop of 
cylindrical radius $r$, $\sigma$ is the ratio of the Poynting flux to the matter 
(kinetic plus rest-mass) energy flux, and $ \Gamma \sigma=-\Omega I/{2\pi\massint c^3}$
is the Poynting flux per unit rest-mass energy flux. 
It is easy to see that the Lorentz factor $\Gamma$
cannot exceed $\enint$.

%fffffffffffffffffffffffffffffffffffffffffffffffffffffffffffffffffff
\begin{figure}[H]
%\includegraphics[width=75mm,angle=-90]{figures/gbu-sig.eps}
%\center
%\centerline
\begin{center}
{\epsfig{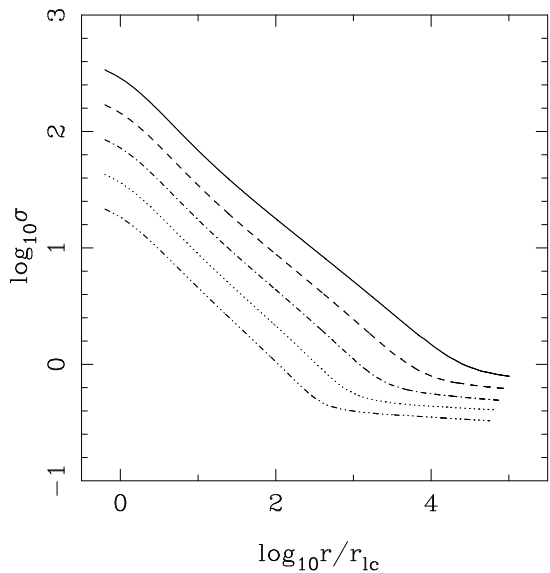}}
\end{center}
\caption{Evolution of $\sigma$ along the magnetic field line $\Psi=0.8\Psi_{max}$ in
models with $\sigma = $ 620 (solid line), 310 (dashed line), 155 (dash-dotted line), 
78 (dotted line) and 39 (dash-triple-dotted line).   
}
\label{sigma-ev-a1}
\end{figure}
%fffffffffffffffffffffffffffffffffffffffffffffffffffffffffffffffffff

%%%%%%%%%%%%%%%%%%%%%%%%%%%%%%%%%%%%%%%%%%%%%%%%%%
\section{Numerical Setup}
%%%%%%%%%%%%%%%%%%%%%%%%%%%%%%%%%%%%%%%%%%%%%%%%%%
\label{simulations}

In the simulation we deal with winds, or unconfined flows, and jets, 
or flows confined with a funnel. In order to avoid complications due 
to numerical diffusion through the jet boundary we use solid funnels 
of paraboloidal shape, $\nonumber z \propto r^a\, ,$
where $z$ and $r$ are the cylindrical coordinates.

%fffffffffffffffffffffffffffffffffffffffffffffffffffffffffffffffffff
\begin{figure}[H]
\begin{center}
 {\epsfig{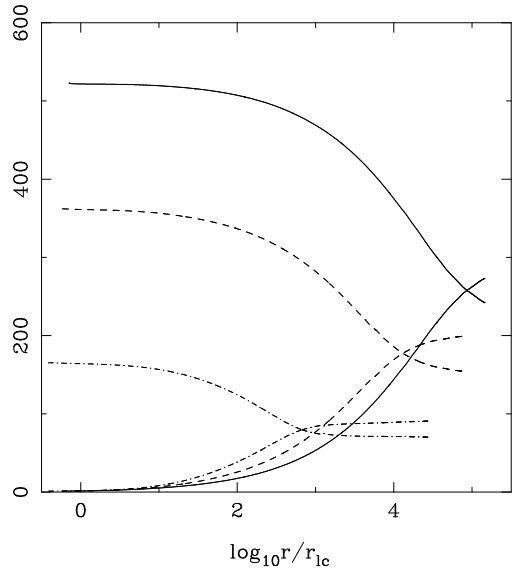}}
\end{center}
\caption{$\Gamma\sigma$ (upper branch)  and $\Gamma$ (lower branch) 
along the magnetic field line $\Psi=0.8\Psi_{max}$ (solid lines), 
along the magnetic field line $\Psi=0.5\Psi_{max}$ (dashed lines), and
along the magnetic field line $\Psi=0.2\Psi_{max}$ (dash-dotted lines) in model $\sigma_0=620$.
}
\label{gamma}
\end{figure}
%fffffffffffffffffffffffffffffffffffffffffffffffffffffffffffffffffff

The initial configuration corresponds to a non-rotating, purely poloidal
magnetic field with approximately constant magnetic pressure across the
funnel. The plasma density within the funnel is set to a small value so
that the outflow generated at the inlet boundary can easily sweep it
away. In order to speed this process up the longitudinal component of velocity
inside the funnel and at the inlet is set to $0.7\, c$.  
In fact, we use the same type of initial and boundary conditions as described 
in our study of lower magnetization jets \cite{kbvk2007}. We constructed a grid of 
models with different funnel power geometry ($a=1,3/2,2$) including unconfined wind, 
different initial magnetization ($\sigma = 10\div600$), and both solid and 
differential rotation at the base.  

%fffffffffffffffffffffffffffffffffffffffffffffffffffffffffffffffffff
%\begin{figure}[H]
%\includegraphics[width=75mm,angle=0]{eqvr_c.eps}
%\center
%\centerline
%\begin{center}
% {\epsfig{figure=figures/barkov-fig2.eps,width=6cm}}
%\end{center}
%\caption{Dependence of ‘equilibrium’ radius $r_{eq}$ from maximal magnetization $\mu$, here 
%$r_{eq}$ is the radius there $\sigma=1$ or half of magnetic energy transferred to kinetic energy of matter.
%}
%\label{eqvr_c}
%\end{figure}
%fffffffffffffffffffffffffffffffffffffffffffffffffffffffffffffffffff

%%%%%%%%%%%%%%%%%%%%%%%%%%%%%%%%%%%%%%%%%%%%%%%%%%%%%%%%%%%%%%%%%
\section{Results}
%%%%%%%%%%%%%%%%%%%%%%%%%%%%%%%%%%%%%%%%%%%%%%%%%%%%%%%%%%%%%%%%%
\label{results}

Figure \ref{sigma-ev-a1} shows the evolution of $\sigma$ along the jet boundary for 
the models with $a=3/2$, solid rotation, and the maximum magnetization at the base  
$\sigma_0=39,78,155,310$ and 620 (for solid rotation $\sigma_0$ increases with r).  
One can see that all solutions 
exhibit transition to the particle dominated regime ($\sigma<1$). 
Prior to reaching the equipartition they are described by the same law 
$\sigma \propto (r/r_{lc})^{-3/5}$ and after this the evolution of $\sigma$ slows down
significantly.  The dependence of `the equipartition` radius $r_{eq}$ on $\sigma_0$ 
can be approximated by 
\begin{equation}
  r_{eq} \simeq 0.079 \sigma_0^{2.1}r_{lc} \mbox{ if } 10<\sigma_0<620.
	\label{req}
\end{equation}
Figure \ref{gamma} confirms similar evolution inside the jets. 
The normal pressure at the jet boundary is close to $p_n\propto R^{-2}$,
where $R$ is the spherical radius. In astrophysical conditions this pressure 
has to be balanced by the pressure of confining medium.

%fffffffffffffffffffffffffffffffffffffffffffffffffffffffffffffffffff
%\begin{figure}[H]
%\includegraphics[width=75mm,angle=-90]{figures/pext.eps}
%\center
%\begin{center}
% \epsfig{figure=figures/barkov-fig3.eps,width=6cm,angle=-90}
%\end{center}

%\centerline{\epsfig{figure=pext.eps,width=6truecm}}
%\caption{Evolution of total pressure along the jet boundary in 
%models A (solid line), B1 (dashed line), C (dash-dotted line), and D (dotted line).
%}
%\label{p-ext}
%\end{figure}
%fffffffffffffffffffffffffffffffffffffffffffffffffffffffffffffffffff

%fffffffffffffffffffffffffffffffffffffffffffffffffffffffffffffffffff
\begin{figure*}
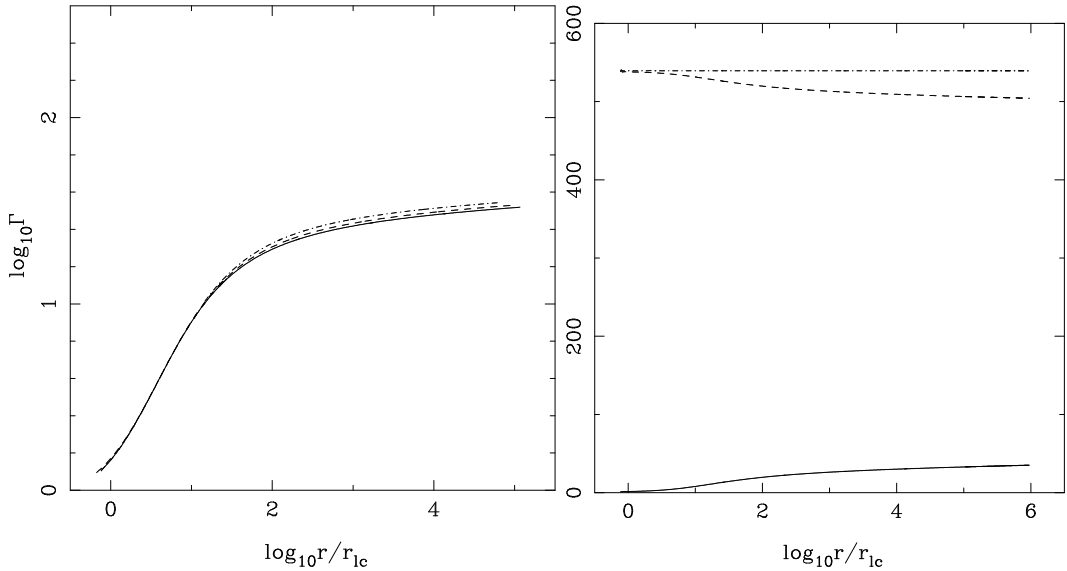

{\epsfig{figure=figures/barkov-fig3a.eps,width=7.5cm,angle=-90}}
{\epsfig{figure=figures/barkov-fig3b.eps,width=7.5cm,angle=-90}}
\caption{Unconfined wind solution. Left panel: 
Lorentz factor along three different magnetic field lines:  
solid line - $\Psi=0.8\Psi_{max}$,
dashed line - $\Psi=0.5\psi_{max}$,
dash-dotted line - $\Psi=0.2\Psi_{max}$. 
Right panel: 
$\Gamma\sigma$  (solid line), $\enint$  (dashed line) and
$\Gamma$ (dash-dotted line) along the magnetic field line with $\Psi=0.8\Psi_{max}$
as a function of cylindrical radius. 
}
\label{gpw}
\end{figure*}
%fffffffffffffffffffffffffffffffffffffffffffffffffffffffffffffffffff

Figure \ref{gpw} shows our results for the unconfined wind.    
One can see that magnetic acceleration in this case is much less effective
in agreement with previous studies \cite{bkr98,bt99}.

%sssssssssssssssssssssssssssssssssssssssssssssss
\section{Application to GRB and AGN Jets and PWN}
%sssssssssssssssssssssssssssssssssssssssssssssss
\label{application}

The initial energy-to-mass flux ratio of jets in our simulations
yields an upper limit on the terminal Lorentz factor $\Gamma_\infty=
\enint $. In order to make further
comparisons of our numerical models with observations we need to
select suitable dimensional scales. The key scale in the problem of
magnetic acceleration is the light cylinder (or the Alfv\'en surface)
radius, $r_{lc}$.  If the jets are launched by a rapidly rotating
black hole in the center of an GRB or AGN then
\begin{eqnarray}
\nonumber r_{lc}\simeq 4 r_g = 2\times 10^{6} ( {M}/{3 \,
M_{\odot}} ) \mbox{ cm}\, ,
\end{eqnarray}
where $r_g\equiv GM/c^2$. In this estimate we assume that the angular
velocity of the magnetic field lines is half of that of a maximally
rotating (rotation parameter $a\simeq1.0$) black hole.  According to the
results shown in equation (\ref{req}), the jets enter the matter-dominated
regime at a cylindrical radius
\begin{eqnarray}
\nonumber r_{eq} \simeq 5.9\times 10^{9} (\mu/150)^{2.1}({M}/{3
M_{\odot}}) \mbox{ cm}\, .
\end{eqnarray}
The corresponding distance from the black hole is
\begin{eqnarray}
\nonumber
   R_{eq} \simeq  5.9\times 10^{10} (\mu/150)^{2.1}({M}/{3
M_{\odot}})(0.1/\Theta_j) \mbox{ cm}\, ,
\end{eqnarray}
where $\Theta_j$ is the jet opening half-angle. Strong shock
waves can appear only if $\sigma <1$. For GRB jets, with $\mu\ge150$ and 
$M\simeq 3M_{\odot}$ this distance is remarkably close to the radius of Wolf-Raye stars.
Thus, in the collapsar scenario the conditions for development of strong fast shocks and 
associated gamma-ray emission are already satisfied when the jets break away from the 
progenitor star. The actual location of shocks, however, also depends on the scale of 
the central engine variability which puts it to larger distances.

Lower Lorentz factor of AGN jets imply lower magnetization parameter, 
$\mu\le30$. Combined with the black hole mass $M\simeq10^{8} M_{\odot}$ this 
gives  $R_{eq} \simeq 2\times 10^{17}$cm which is similar to the size of the 
``blazar zone'' inferred from the observations. 
Thus, the slow nature of magnetic acceleration 
of relativistic jets is in very good agreement
with the observational constraints.

%{\bf What difference? Describe.}
The velocity profile of jets with a solid body rotation at their base 
shows strong tangential discontinuity at the jet boundary which favors 
the photon-breading mechanism of gamma-ray emission from AGN jets 
\cite{agr06,der03,stn07}. On the contrary, the gradual decline of Lorentz 
factor towards the jet boundary which we see in the models with 
differential rotation at the base significantly reduces its efficiency.
This implies that gamma-gay observations can be used to determine whether 
the AGN jets originate directly from the black hole magnetospheres or from the 
magnetospheres of their accretion discs.

%GRB and AGN jets could be confined by a variety of
%forces, including, for example, the thermal pressure of an ambient gas
%distribution, the ram pressure of a wind from the outer regions of the
%nuclear disk, the ram pressure of the accretion stellar envelop, 
%and the stress of a magnetic field anchored in the disk
%(and possibly embedded in a disk outflow). Although none of the
%curves on Fig. \ref{p-ext} is an exact power law of the form $p_{\rm ext} \propto
%R^{-\alpha}$, it is nevertheless informative to calculate mean
%power-law indices. We find $\alpha \approx 4.6\div2.4$, 2 and 1.85 for
%models A, B1 and C, respectively. 

Like other researches \cite{bkr98,bt99} we find that magnetic acceleration of 
highly magnetized unconfined winds is less ineffective. This is a rather uncomfortable 
conclusion since the successful MHD model of Pulsar Wind Nebulae \cite{kc84,kl04,dab04} 
requires the pulsar wind to be particle dominated at the location of its termination 
shock. The two possible solutions to this problem suggested so far utilize the alternating 
structure of magnetic field in the wind from oblique rotators (`striped wind', \cite{mich82}). 
Firstly, the alternating magnetic field can dissipate before reaching the termination 
shock, e.g. \cite{ar07}, with eventual convertion of the released heat into the bulk 
motion energy. Secondly, it can dissipate inside the shock layer of the termination 
shock itself\cite{pl07}. In this case one cannot directly use the standard shock 
equations in order to determine the downstream state.

%%%%%%%%%%%%%%%%%%%%%%%%%%%%%%%%%%%%%%%%%%%%%%%%%%%%%%%%%%%%%%%%%
\section{Conclusion}
%%%%%%%%%%%%%%%%%%%%%%%%%%%%%%%%%%%%%%%%%%%%%%%%%%%%%%%%%%%%%%%%%
\label{conclusion}

In validating the basic features of the simplified semi-analytic
solutions, our numerical results go a long way toward establishing an
``MHD acceleration and collimation paradigm'' for relativistic
astrophysical jets. In particular, they demonstrate that even jets with 
extremely high initial magnetization can be effectively accelerated via the 
ideal magnetic mechanism with more than a half of the Poynting flux converted into 
the bulk motion energy of the flows. The highest Lorentz factor reached in the 
simulations is $\Gamma=300$ which is well within the range deduced for GRB jets. 
The slow character of magnetic acceleration allows dissipationless energy 
transport over large distances, the property which is deduced from observations and 
which is rather difficult to explain in other models of jet generation.

Interested reader can find more details about the method of our simulations 
and results for AGN jets in \cite{kbvk2007}.

\end{multicols}

%\bibliographystyle{ws-procs11x85}
%\bibliography{ws-pro-sample}

\begin{thebibliography}{99}

\bibitem{C07}
{Cohen} M.~H. {\em et al.}, ApJ, {\bf 658}, 232,  (2007).



\bibitem{J05}
{Jorstad} S.~G. {\em et al.}, \aj, {\bf 130}, 1418, (2005).

\bibitem{J01}
{Jorstad} S.~G. {\em et al.}, \apjs, {\bf 134}, 181,  (2001).


\bibitem{S05}
Sikora M. {\em et al.}, ApJ, {\bf 625}, 72,  (2005).


\bibitem{rud75}
Rederman M., {\em Ann. N.Y. Acad. Sci}, {\bf 262}, 164,  (1975).

\bibitem{kp91}
Krolik J.H. {\em et al.}, \apj, {\bf 373}, 277,  (1991).

\bibitem{pir99}
Piran T., {\em Phys. Rep.}, {\bf 314}, 575,  (1999).

\bibitem{tay04}
Taylor G.B. {\em et al.}, \apj, {\bf 609}, L1,  (2004).

\bibitem{BZ77}
{Blandford} R.~D.{\em et al.}, \mnras, {\bf 179}, 433, (1977).


\bibitem{L76}
Lovelace R.~V.~E., Nature, {\bf 262}, 649, (1976).




\bibitem{BBR84}   
{Begelman} M.~C. {\em et al.}, {\em Reviews of Modern Physics}, {\bf 56}, {255}, {(1984).}


\bibitem{BR74} 
{Blandford}  R.~D. {\em et al.},  {\em MNRAS} , {\bf 169}, {395}, {(1974).}

\bibitem{VK03a}
{Vlahakis} N. {\em et al.}, \apj, {\bf 596}, 1080,  (2003a).

\bibitem{VK03b}
{Vlahakis} N. {\em et al.}, \apj, {\bf 596}, 1104,  (2003b).


\bibitem{bn06}
{Beskin} V.S. {\em et al.}, {\em MNRAS}, {\bf 367}, 275,  (2006).

\bibitem{kbvk2007}
{Komissarov} S.~S. {\em et al.}, \mnras, {\bf 380}, 51,  (2007).

\bibitem{bkr98}
{Beskin} V.S. {\em et al.}, {\em MNRAS}, {\bf 299}, 341,  (1998).

\bibitem{bt99}
{Bogovalov} S. {\em et al.}, {\em MNRAS}, {\bf 305}, 211,  (1999).

\bibitem{agr06}
{Aharonian} F. {\em et al.}, {\em Science}, {\bf 314}, 1424,  (2006).

\bibitem{der03}
{Derishev} E.V. {\em et al.}, {\em Physical Review D}, {\bf 68}, 043003,  (2003).

\bibitem{stn07}
{Stern} B.E. {\em et al.}, {\em MNRAS.tmp.1194S},  (2007).

\bibitem{kc84}
{Kenel} C.F. {\em et al.}, {\em ApJ}, {\bf 283}, 694, (1984).

\bibitem{kl04}
{Komissarov} S.~S. {\em et al.}, \mnras, {\bf 349}, 779,  (2004).

\bibitem{dab04}
{Del Zana} L. {\em et al.}, {\em A\&A}, {\bf 421}, 1063,  (2004).


\bibitem{mich82}
{Michel} F.~C., {\em Reviews of Modern Physics}, {\bf 54}, 1,  (1982).

\bibitem{ar07}
{Arons} J., {\em astro-ph/0710.1326B}, (2007).

\bibitem{pl07}
{P\'{e}tri} J. {\em et al.}, {\em A\&A}, {\bf 473}, 683,  (2007).


\end{thebibliography}
%\end{multicols}
\end{document}